\documentclass[letterpaper,english,apl,preprint,superscriptaddress]{revtex4}
\usepackage[T1]{fontenc}
\usepackage[latin9]{inputenc}
\usepackage{units}
\usepackage{amssymb}
\usepackage{graphicx}

\makeatletter

\@ifundefined{textcolor}{}
{%
 \definecolor{BLACK}{gray}{0}
 \definecolor{WHITE}{gray}{1}
 \definecolor{RED}{rgb}{1,0,0}
 \definecolor{GREEN}{rgb}{0,1,0}
 \definecolor{BLUE}{rgb}{0,0,1}
 \definecolor{CYAN}{cmyk}{1,0,0,0}
 \definecolor{MAGENTA}{cmyk}{0,1,0,0}
 \definecolor{YELLOW}{cmyk}{0,0,1,0}
 }

\usepackage{gensymb}

\makeatother

\usepackage{babel}
\begin{document}

\title{Quantum Hall effect on centimeter scale chemical vapor deposited
graphene films}

\author{Tian Shen}

\email{tshen@nist.gov}

\affiliation{Department of Physics, Purdue University, West Lafayette, IN, 47907}

\affiliation{Physical Measurement Laboratory, National Institute of Standards
and Technology, Gaithersburg, MD, 20899}

\author{Wei Wu}

\affiliation{Center for Advanced Materials and ECE, University of Houston, Houston,
TX, 77204}

\author{Qingkai Yu}

\affiliation{Center for Advanced Materials and ECE, University of Houston, Houston,
TX, 77204}

\author{Curt A Richter}

\affiliation{Physical Measurement Laboratory, National Institute of Standards
and Technology, Gaithersburg, MD, 20899}

\author{Randolph Elmquist}

\affiliation{Physical Measurement Laboratory, National Institute of Standards
and Technology, Gaithersburg, MD, 20899}

\author{David Newell}

\affiliation{Physical Measurement Laboratory, National Institute of Standards
and Technology, Gaithersburg, MD, 20899}

\author{Yong P. Chen}

\affiliation{Department of Physics, Purdue University, West Lafayette, IN, 47907}

\affiliation{Birck Nanotechnology Center, Purdue University, West Lafayette, IN
47907}

\affiliation{School of Electrical and Computer Engineering, Purdue University,
West Lafayette, IN 47907}
\begin{abstract}
We report observations of well developed half integer quantum Hall
effect (QHE) on mono layer graphene films of 7 mm $\times$ 7 mm in
size. The graphene films are grown by chemical vapor deposition (CVD)
on copper, then transferred to SiO$_{2}$/Si substrates, with typical
carrier mobilities $\approx$4000 cm$^{2}$/Vs. The large size graphene
with excellent quality and electronic homogeneity demonstrated in
this work is promising for graphene-based quantum Hall resistance
standards, and can also facilitate a wide range of experiments on
quantum Hall physics of graphene and practical applications exploiting
the exceptional properties of graphene.
\end{abstract}

\date{\today}

\maketitle
Graphene, a single sheet of carbon atoms tightly packed into a two-dimensional
(2D) hexagonal lattice, has been considered as a promising candidate
for the next generation high-speed electronic devices due to its extraordinary
electronic properties, such as a carrier mobility exceeding $10\ 000$
cm$^{2}$/Vs and an electron velocity of $\approx10^{8}$ cm/s at
room temperature.\citep{Geim2007} Monolayer graphene's massless fermion
nature gives rise to a characteristic {}``half-integer'' quantum
Hall effect (QHE). The associated Landau level (LL) separation, $\Delta(E)=(\sqrt{n+1}-\sqrt{n})v_{F}\sqrt{2eB\hbar}$,
is anomalously large for even moderate magnetic fields.\citep{Zhang2005,Novoselov2005a}
For example, at a magnetic field of 10 T, the LL gap between the $n$=0
and $n$=$\pm$1 LL is $\approx$1300 K in graphene compared to $\approx$200
K in GaAs heterostructures, enabling the observation of QHE at high
temperatures,\citep{Novoselov2007} and raising exciting possibilities
for future quantum Hall resistance metrology based on graphene.\citep{Giesbers2008,Tzalenchuk2010}
Graphene flakes, typically only a few microns in size, as  produced
by mechanical exfoliation onto SiO$_{2}$ \citep{Novoselov2004} are
considered to be too small for resistance metrology applications.
The recent demonstration of half-integer quantum Hall effect in monolayer
graphene epitaxially grown on both Si face\citep{Shen2009,Tzalenchuk2010,Jobst2010}
and C face\citep{Wu2009} of SiC, and grown by chemical vapor deposition
(CVD) on Ni\citep{Kim2009a} and Cu\citep{Cao2010,Bae2010}, suggests
that graphene produced by such more scalable approaches can have advantages
for metrology applications. However, due to charge inhomogeneity\citep{Martin2008},
QHE has so far only been observed in samples of size up to $\approx$100
$\mu$m$\times$100 $\mu$m. In this letter, we demonstrate CVD graphene
grown on Cu with size approaching centimeter scale possessing excellent
electronic properties as evidenced by well-developed half-integer
QHE. Our results are important not only for the development of graphene
based quantum Hall resistance standards, but also for other practical
applications of graphene that desire large scale samples with excellent
electronic quality and uniformity, such as integrated circuits and
transparent conducting electrodes.\citep{Kim2009a} Such high-quality
and large-scale graphene films will also facilitate a wide range of
experiments, such as optical studies and scanning tunneling microscope
(STM) studies of graphene's QHE and other electronic properties.

\begin{figure}
\includegraphics[width=1\columnwidth]{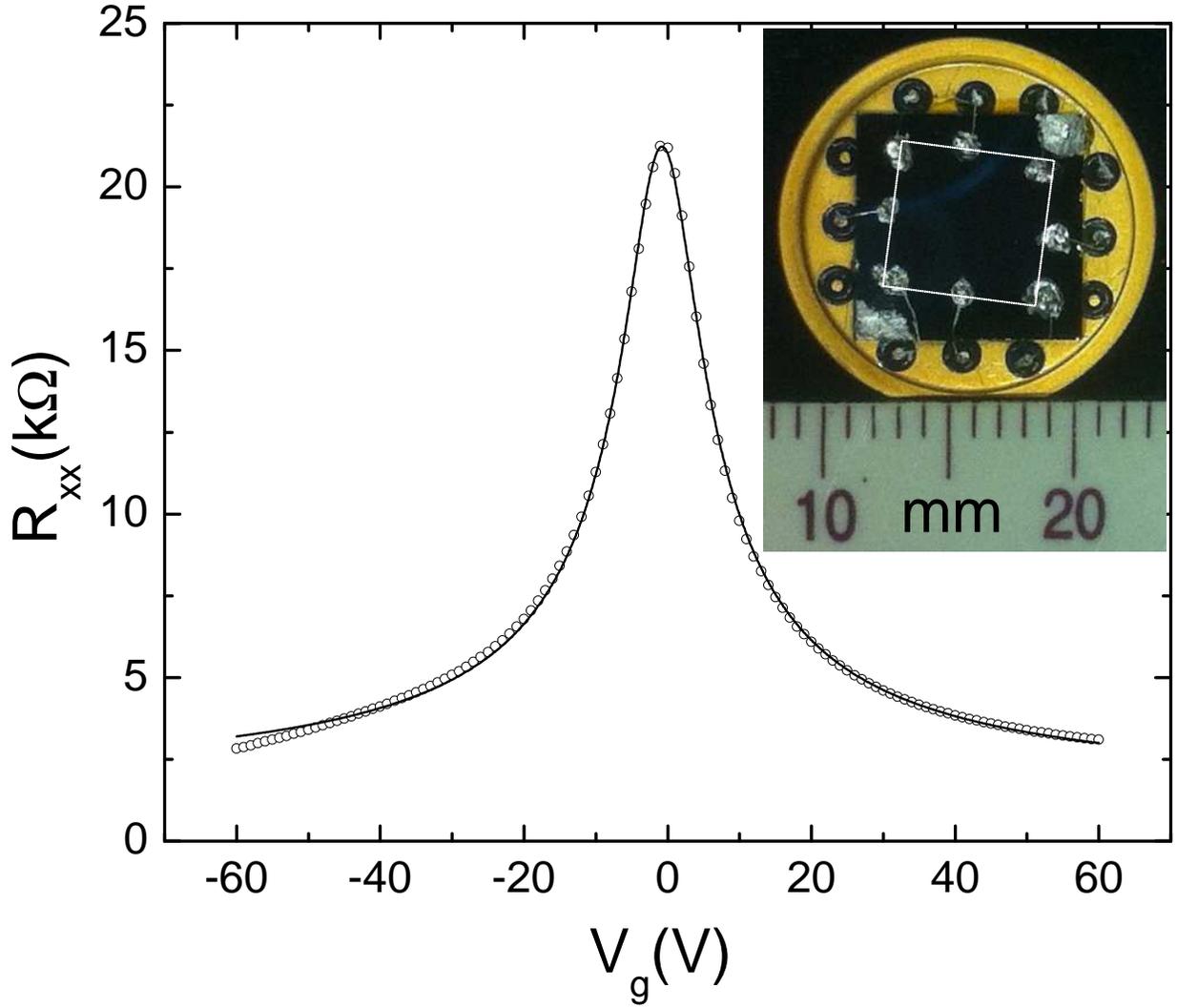}

\caption{DC four-terminal longitudinal resistance $R_{xx}$ (hollow circle)
measured as a function of gate voltage $V_{g}$ at 1.7 K along with
modeling fit (solid line). Insert: Picture of the 7 mm$\times$7 mm
graphene device on a SiO$_{2}$/Si substrate. The dashed square is
a guide to the eye indicating the outline of the graphene film.}

\label{Fig:SamplePicture}
\end{figure}

The graphene samples studied here are synthesized by CVD of CH$_{4}$
on Cu foils (25 $\mu$m thick, 99.8 \%) in a quartz tube furnace at
ambient pressure.\citep{Cao2010,Wu2010} Before graphene deposition,
Cu foils are annealed in Ar and H$_{2}$ at 1050 $\celsius$ for 30
min to clean the Cu surface and increase the Cu grain size. Graphene
growth is then carried out under 15 parts-per-million (ppm) CH$_{4}$
(balanced in Ar and H$_{2}$) at 1050 $\celsius$ for 25 min. To transfer
graphene to SiO$_{2}$/Si substrates, the as-synthesized samples are
spin-coated by polymethyl methacrylate (PMMA) and placed in an aqueous
solution of iron nitrate to etch away the Cu substrate. Afterwards,
the graphene with PMMA coating is scooped out from the solution directly
onto the transfer substrate. After carefully rinsing in acetone and
isopropanol to remove the PMMA residues induced in the transfer process,
indium contacts both to graphene and backside silicon are made to
form a Van der Pauw device, as shown in Fig. \ref{Fig:SamplePicture}
insert. The sample is then immediately cooled down in a variable temperature
$^{4}$He cryostat (1.6 K to 300 K) to minimize the exposure to atmosphere,
which introduce hole doping and increase disorder in the graphene
films.\citep{Ryu2010} Four-point magneto-transport measurements are
performed in magnetic fields up to 14 T using both the low frequency
ac lock-in technique and current reversed dc technique with a source-drain
(SD) input current of 100 nA for fast or precise characterizing of
the device performance. We perform the dc resistance measurements
using a dc current source and three $8\nicefrac{1}{2}$ digit multimeters,
two for $V_{xx}$, $V_{xy}$ and one for $V_{cal}$, measured across
a 10 k$\Omega$ resistance standard (in series of the sample) whose
value is known better than 1 ppm. The measured values of $V_{cal}$
are used for cross-checking the current and calculating  of $R_{xx}$
and $R_{xy}$. The carrier density is tuned by a back gate voltage
$V_{g}$ applied between the highly doped Si substrate and the graphene,
with the 300 nm SiO$_{2}$ as the gate dielectric. Measurements on
two such devices yield similar results, however data for only one
of them are presented below.

\begin{figure}
\includegraphics[width=1\columnwidth]{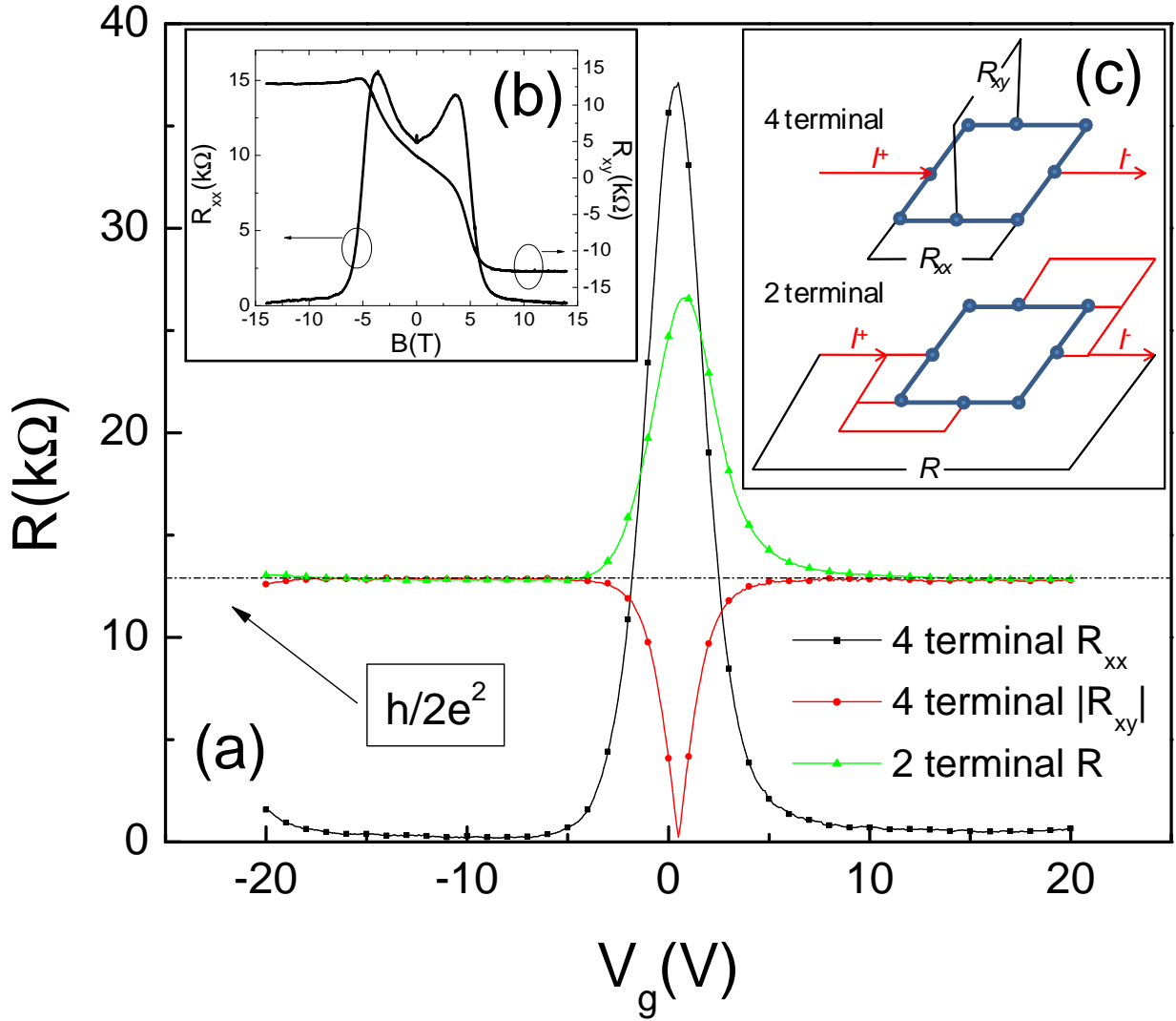}

\caption{(a) DC four-terminal longitudinal resistance $R{}_{xx}$, absolute
value of Hall resistance $|R{}_{xy}|$ and two-terminal resistance
$R$ as a function of $V_{g}$ measured at 1.7 K and a magnetic field
of 14 T. The resistances are (anti)symmetrized between positive and
negative magnetic fields. The horizontal dashed line is the $h/2e^{2}$
conventional value as assigned in 1990.\citep{Taylor1989} (b) $R_{xx}$
and $R_{xy}$ as a function of magnetic field at 1.7 K and $V_{g}$=-9
V. (c) Schematic circuit diagram for four terminal and two terminal
triple-series measurements. The solid dots represent the indium contacts
as shown in Fig. \ref{Fig:SamplePicture} insert. }

\label{Fig:DC measurement}
\end{figure}

Figure \ref{Fig:SamplePicture} shows the dc four terminal longitudinal
resistance $R_{xx}$, configured as in Fig. \ref{Fig:DC measurement}
(c), measured at 1.7 K and zero magnetic field with $V_{g}$ from
-60 V to 60 V. It shows ambipolar field effect with resistance modulation
ratio greater than seven. The Dirac point, $V_{Dirac}$, in the device
is situated at -0.7 V, indicating a very low extrinsic charge doping
level. This is expected since no fabrication is performed on this
device, and the contamination introduced during common fabrication
processes is minimized. In our sample, the accuracy of the simple
Van der Pauw method to extract the resistivity (thus the mobility)
may be limited  due to rough edges, large contact sizes and voids
inside the film that change the current path. Therefore, we have used
a self-consistent approach to fit the field effect curve (see %
\footnote{An empirical fitting to the field effect curve (Fig. \ref{Fig:SamplePicture})
according to \citet{Kim2009} is used to calculate a geometry-independent
residual carrier density $n_{0}=3.7\times10^{11}$ cm$^{-2}$. From
$n_{0}$ and a theoretical model on graphene transport in the diffusive
region by \citet{Adam2007}, one can extract that the impurity concentration
$n_{imp}=(1.4\pm0.2)\times10^{12}$ cm$^{-2}$ and a field effect
mobility $\mu=(3600\pm600)$ cm$^{2}$/Vs assuming a typical distance
of impurities from the graphene plane, $d=(1.0\pm0.2)$ nm. This gives
a geometrical factor (ratio between $R_{xx}$ and resistivity $\rho_{xx}$)
of $4.2\pm0.7$ by comparing to the empirical fitting\citep{Kim2009}
of the field effect curve. %
}), and extracted both a geometrical factor of $4.2\pm0.7$ and a field
effect mobility of $(3600\pm600)$ cm$^{2}$/Vs. A Hall mobility of
$(4000\pm700)$ cm$^{2}$/Vs at $V_{g}$= -30 V is obtained by applying
the same geometrical factor, consistent with the field effect mobility.

To further characterize the film quality and characteristic quantum
Hall effect, we have measured $R_{xx}$ and $R_{xy}$ versus $B$
(perpendicular magnetic field)  at fixed $V_{g}=-9$ V as shown in
Fig. \ref{Fig:DC measurement} (b). The $i=\pm2$ quantum Hall plateaus
are well resolved. The higher LL plateaus are still not visible, due
to disorder induced LL broadening. Below we focus on measurements
by sweeping $V_{g}$ at the highest available magnetic field, where
better quality quantum Hall states are observed. The dc measurements
are carried out in both four-terminal and two-terminal configuration
as a function of $V_{g}$ at 1.7 K with $B=14$ T as shown in Fig.
\ref{Fig:DC measurement} (a). The resistances are (anti)symmetrized
between positive and negative magnetic field orientations to compensate
for the imperfect geometry. The longitudinal resistance $R_{xx}$
shows low minimum values indicating quantum Hall states  in the ranges
of -18 V to -5 V for holes and 10 V to 20 V for electrons. The Hall
resistance $R_{xy}$ shows well developed  quantum Hall plateaus at
the corresponding $R_{xx}$ minima. The sharp $i=2$ to $i=-2$ quantum
Hall plateau to plateau transition happens within a gate voltage range
of $(1.5\pm0.4)$ V (measured between two extrema in the derivative
$d\sigma_{xx}/dV_{g}$ for $n=0$ LL), comparable with the transition
for exfoliated graphene.\citep{Giesbers2009} This also confirms that
the inhomogeneity in our large graphene film is small, as large inhomogeneity
will tend to broaden the transition and even destroy the quantum Hall
effect. The observed quantum Hall resistances are not exact, showing
a residual $R_{xx}$ of $\approx200\ \Omega$ for holes ($i=-2$ state)
and $\approx500\ \Omega$ ($i=+2$ state) for electrons. $R_{xy}$
also deviates from $h/2e^{2}$ (the dashed horizontal line) by $\approx50\ \Omega$
for holes and $\approx100\ \Omega$ for electrons, qualitatively consistent
 with the theory that the absolute error in the quantization of $\rho_{xy}$
(=$R_{xy}$) can be related to a finite resistivity $\rho_{xx}$ (indicating
dissipation) as $\Delta\rho_{xy}=-s\rho_{xx}$, where s is on the
order of unity.\citep{Furlan1998} While the four-terminal resistance
measurement is widely used to eliminate the effect of contacts and
wires, two terminal triple-series connections proposed by \citet{Delahaye1993,Delahaye1995}
can also be used in both dc and ac bridges for high precision measurements
of QHE. Assuming all of the contact and wire resistance values $R_{c},R_{w}\ll R_{H}=h/2e^{2}$,
then from the dc equivalent circuit model, \citep{Jeffery1995} the
resulting two-terminal resistance for a perfectly quantized $i=\pm2$
plateau is:
\begin{equation}
R=R_{H}\left(1+2\left(\frac{R_{c}+R_{w}}{R_{H}}\right)^{3}\right).
\end{equation}
Considering a few ohms of resistance presented in the wire, then $R_{c}<10\ \Omega$
is generally desired to achieve the precision of a few parts per billion
when using $R$ to measure the QHE. This is an experimental challenge
for micron-sized graphene devices since the best normalized contact
resistance so far is still $\approx500\ \Omega\mu$m.\citep{Russo2010}
Large contact resistance also introduces noise on the voltage probes,
and leads to local heating at the current contacts thereby limiting
the maximum breakdown current of QHE. The scalability of CVD graphene
makes it possible to use large area contacts to reduce contact resistances.
In our device, the contact size is $\approx$1 mm, thus $R_{c}$ can
easily fall below 10 $\Omega$ assuming a reasonable normalized contact
resistance, i.e. $\approx5$ k$\Omega\mu$m for indium and clean graphene
surface. While the precise determination of the contact resistance
from our measured QHE plateau is difficult due to a non-vanishing
$R_{xx}$, the measured two terminal $R$ agrees well with four terminal
$R_{xy}$ at the center of plateau, indicating that the contact resistance
is likely much smaller than the residual $R_{xx}$ and does not play
a significant role here.

\begin{figure}
\includegraphics[width=1\columnwidth]{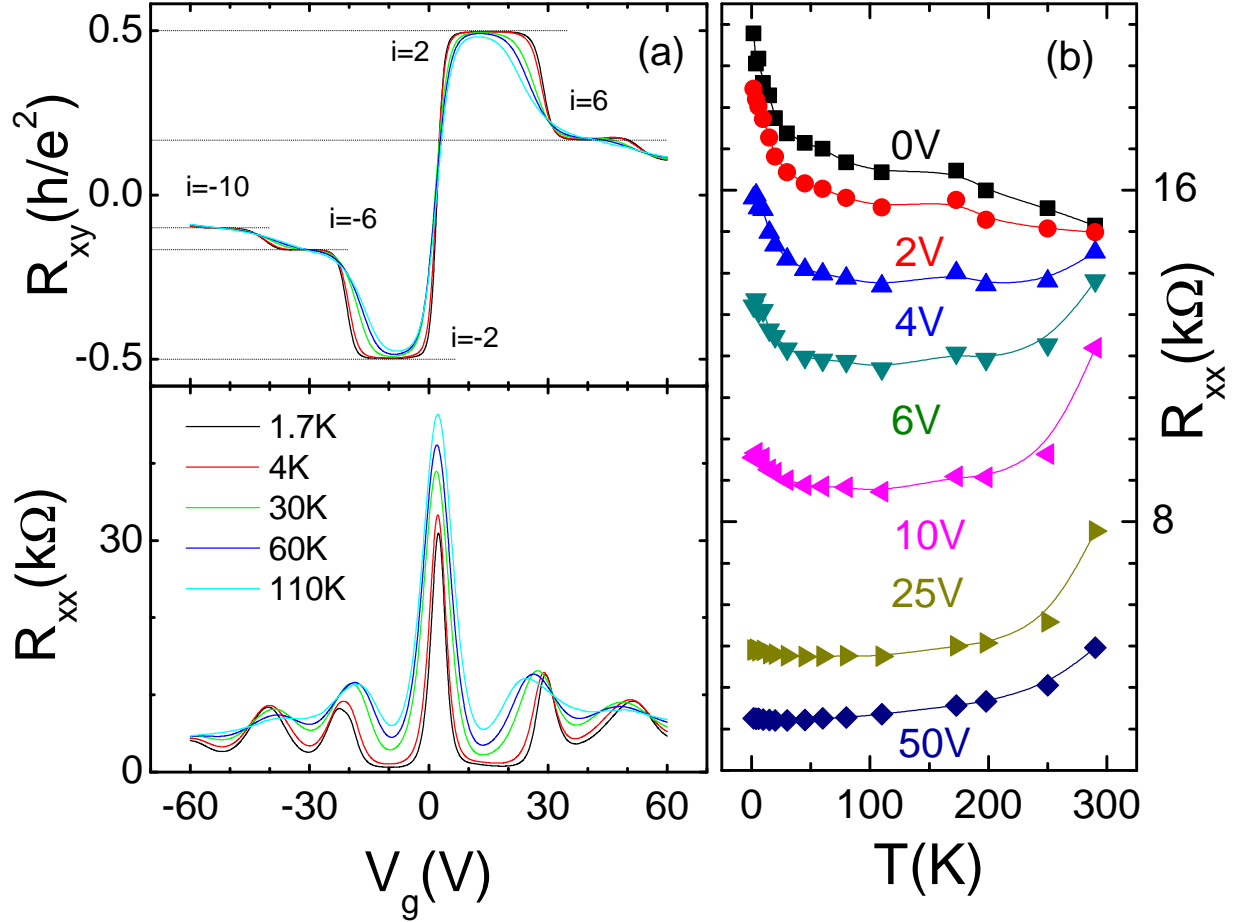}

\caption{(a) Hall resistance $R_{xy}$ and longitudinal resistance $R_{xx}$
as a function of $V_{g}$ at B=14 T for different temperatures. The
resistances are (anti)symmetrized between the positive and negative
magnetic field orientations. The horizontal dotted lines indicate
the $i=\pm2$, $\pm6$ and $-10$ quantum Hall plateau conventional
values as assigned in 1990.\citep{Taylor1989} (b) $R_{xx}(T)$ at
various $\Delta V_{g}$ at $B$=0 T. The symbols are experimental
data and the lines are guides to the eye.}

\label{Fig:TempDependency}
\end{figure}

To investigate the reliability of this device, the sample is warmed
up and exposed to atmosphere for 24 h and then cooled back down again.
An ac lock-in technique is used for fast characterization. $R_{xx}$
and $R_{xy}$ are measured as a function of $V_{g}$ at B=$\pm$14
T for different temperatures, and (anti)symmetrized as before (the
gate sweep is performed over a wider range than that in Fig. \ref{Fig:DC measurement},
therefore revealing more quantum Hall states). We observe an increase
in $V_{Dirac}$ to 2.4 V and the QHE residual longitudinal resistance
$R_{xx}$ to $\approx600\ \Omega$ ($\approx750\ \Omega$) for holes
(electrons) at 1.7 K, indicates an increase of hole doping (and possibly
disorder) from exposure to the atmosphere. A charge-neutral passivation
layer similar to those reported for graphene on SiC\citep{Lara-Avila2011}
may be employed in future to preserve the graphene quality in ambient
environment. In any case, as  shown in Fig. \ref{Fig:TempDependency}(a),
the $i=\pm2$ quantum Hall plateaus are still pronounced up to 60
K, showing the great potential as a quantum Hall resistance standard
that can be used at much higher temperatures than those using GaAs
heterostructures.

We have also characterized the zero magnetic field temperature dependent
$R_{xx}$ for this device at different $\Delta V_{g}$ relative to
the Dirac point as shown in Fig. \ref{Fig:TempDependency} (b), where
$\Delta V_{g}=V_{g}-V_{Dirac}$, to get a qualitative indication of
the disorder in the graphene. The resistance at the Dirac point increased
by about 30 \% with decreasing temperature between 300 K and 1.7 K.
For diffusive transport in the {}``dirty limit'', this temperature
dependence is dominated by activation across potential barriers in
inhomogeneous puddles.\citep{Adam2010} The rapid increase of resistance
at temperatures less than 20 K is due to the weak localization. The
temperature dependence becomes non-monotonic near $\Delta V_{g}\approx4$
V where the carrier density $n_{s}=2.9\times10^{11}$ cm$^{-2}$,
similar to results reported in Ref. \citep{Heo2011}, showing that
there still exist a fair amount of impurity scattering compared with
exfoliated graphene or clean CVD graphene of very small size. At the
high carrier density limit ($\Delta V_{g}=50$ V) where the puddle
effect is suppressed, $R_{xx}(T)$ exhibits metallic behavior, indicating
scattering dominated by surface phonons. A better transfer technique
is desired to best preserve the quality of as-grown CVD graphene .

In summary, we have studied the magneto-transport of large scale mono
layer graphene grown by CVD on Cu, then transferred to SiO$_{2}$/Si.
In a 7 mm $\times$ 7 mm Van der Pauw geometry, these devices show
half integer QHE at temperatures up to 110 K at B=14 T, with a carrier
mobility near 4000 cm$^{2}$/Vs. Such CVD graphene brings promising
opportunities for graphene based  integrated circuits, transparent
electronics, quantum Hall resistance metrology, as well as optical
and STM studies due to its exceptional electronic properties kept
even at very large scale.
\begin{acknowledgments}
Y.P. C. acknowledges the support of NIST MSE grant 60NANB9D9175. The
authors would like to thank Shaffique Adam for discussions.
\end{acknowledgments}
\bibliographystyle{apsrev}

\end{document}